\documentclass[twocolumn,showpacs,pre,a4paper]{revtex4}
\usepackage{amssymb}
\usepackage{amsfonts}
\usepackage{amsmath}
\usepackage{graphicx}
\usepackage{dcolumn}
\usepackage{bm}

\begin{document}

\title{Asymptotically pivotal statistic for surrogate testing with extended hypothesis}

\author{Xiaodong Luo$^\dag$}\email{luox@maths.ox.ac.uk}
\author{Jie Zhang$^\ddag$}
\author{Junfeng Sun$^\ddag$}
\author{Michael Small$^\ddag$}
\author{Irene Moroz $^\dag$}

\affiliation{$^\dag$Oxford Centre for Industrial and Applied
Mathematics, Mathematical Institute, University of Oxford, Oxford,
UK\\
$^\ddag$Department of Electronic and Information Engineering, Hong
Kong Polytechnic University, Hom Hung, Hong Kong}

\date{\today }

\begin{abstract}
The method of surrogate data provides a framework for testing
observed data against a hierarchy of alternative hypotheses. The aim
of applying this method is to exclude the possibility that the data
are consistent with simple linear explanations before seeking
complex nonlinear causes. However, in recent time the method has
attracted considerable criticism, largely as a result of ambiguity
about the formation of the underlying null hypotheses, or about the
power of the chosen statistic. In this communication we show that by
employing a special family of ranks statistics these problems can be
avoided and the method of surrogate data placed of a firm
statistical foundation.
\end{abstract}

\pacs{05.45.Tp, 02.50.-r, 05.10.-a}

\maketitle

Since the framework of surrogate testing \cite{Theiler
testing,Theiler constrained} was proposed, it rapidly became a
popular method to distinguish linear stochastic noise from nonlinear
irregular time series, and was widely applied in many fields
\cite{Nichols detecting,oh statistical,small deterministic} albeit
its intrinsic systematic problems, among which two of the most
noticeable ones are: $\left(1\right)$ ``Wraparound artifact''
\cite{Galka topics,small applied,Theiler detecting}, that is,
adopting the Fourier transform to generate surrogates will result in
artifact circular autocorrelation function; $\left(2\right)$
Inconsistency for the surrogates to represent the null hypothesis
\cite{Kugiumtzis test,Kugiumtzis on}.

A few remedies were proposed to deal with the problems of
$\left(1\right)$ and $\left(2\right)$. For example, for problem
$\left(1\right)$, to avoid the Wraparound artifact, the idea of
limited phase randomization was suggested \cite{Galka
topics,Nakamura trend}. A more elaborate solution (but more
computation-intensive meanwhile) was to adopt the method of
simulated annealing \cite{Schreiber constrained}. For problem
$\left(2\right)$, to decrease the inconsistency between the
surrogates and the null hypothesis, the combination of linear and
nonlinear discriminating statistics was recommended \cite{Kugiumtzis
test}.

A closer examination on problem $\left(1\right)$ and
$\left(2\right)$ reveals that, both problems are mainly ascribed to
the adoption of the Fourier transform. Thus by discarding this
technique one may circumvent those problems as well. As an example,
a new algorithm was recently devised based on the linear
superposition principle \cite{luo surrogate,Nakamura testing}.

This work is going to expose another important, yet less discussed
problem. As pointed out in \cite{Theiler constrained}, a pivotal
discriminating statistic could increase the accuracy of surrogate
testing. Furthermore, with a pivotal statistic, the generated
surrogates need not be constrained realizations, this therefore
presents another possible strategy to relieve the problem of
inconsistency between surrogates and the corresponding hypothesis.
Here, a ``pivotal statistic'' refers to the statistical measure that
has the same distribution for all of the processes consistent with a
composite null hypothesis \cite{Galka topics,small applied,Theiler
constrained}, where a composite null hypothesis means that the set
of the relevant processes is not singleton. As an example, the
correlation dimension was shown to be a pivotal test statistic for
linear Gaussian noise \cite{small correlation}.

Our focus is to derive a test statistic, namely the Wilcoxon signed
rank statistic, which will be demonstrated to be pivotal for a
family of linear stationary noise, including the linear stationary
Gaussian noise considered in \cite{Theiler testing,Theiler
constrained}, and asymptotically pivotal for the summation over
realizations of linear stationary noise otherwise. Moreover, we will
also show that, under our hypothesis (to be specified later), in
general the Wilcoxon statistic will have a known asymptotical
distribution (and even exact in certain situations), which is a
considerable advantage for statistical inference in surrogate
testing \cite{luo exact}, in contract to the popular statistics in
the literature (including the correlation dimension), whose
distributions are often analytically untraceable due to the
complication in their computations.

In this work we confine our discussion to stationary time series
as well \cite{Theiler testing,Theiler detecting,Theiler
constrained}. Our \emph{null hypothesis} is to assume that the
time series $\{x_i\}$ under study is from a general $p$-th order
linear autoregressive process, i.e.,
\begin{equation}
x_{i}=a_{0}+\sum\nolimits_{j=1}^{p}a_{j}x_{i-j}+\epsilon
_{i}\text{,} \label{AR_P}
\end{equation}
with unknown coefficients $\{a_i\}_{i=0}^p$, regressive order $p$
and innovation term $\epsilon _{i}$. In general, based on the
Wold's decomposition \cite{pourahmadi foundations} of a stationary
stochastic time series, the innovation term in Eq. (\ref{AR_P}) is
shown to be white. In particular, if $\epsilon _{i}$ is assumed to
follow a Gaussian distribution, then it returns to the scenario
considered in \cite{Theiler testing,Theiler detecting,Theiler
constrained}.

To derive our test statistic, let us first consider a specific
family of the innovation terms. We call innovation terms $\{\epsilon
_{i}\}$ jointly symmetric if there exists a constant $\mu$ such that
$\{\epsilon _{i}-\mu\}$ and $\{\mu-\epsilon _{i}\}$ follow the same
joint distribution, i.e, their probability density functions satisfy
that $f(\epsilon _{1}-\mu ,\epsilon _{2}-\mu ,...)=f(\mu -\epsilon
_{1},\mu -\epsilon _{2},...)$ \cite{dufour nonparametric}. For
instance, innovation terms with normal or uniform distributions are
jointly symmetric.

For a linear stochastic process $\{x_i\}$ with jointly symmetric
innovation terms, we use an ordinary least square (OLS) predictor
\begin{equation}
\hat{x}_{i}^{k}=a_{i,0}+\sum\nolimits_{j=1}^{p^{\prime}}a_{i,j}\hat{x}_{i}^{k-j}
\label{OLS}
\end{equation}
to obtain the $k$-step ahead prediction value $\hat{x}_{i}^{k}$
given the historical record $\{x_{i},x_{i-1},x_{i-2},...\}$ up to
time index $i$, where $\{a_{i,j}:j=0,1,...,p^{\prime}\}$ are the
coefficients estimated based on the history (up to time index $i$)
with the fitting order $p^{\prime}$. Consequently, we have the
corresponding prediction error
$\hat{e}_{i}^{k}=x_{i+k}-\hat{x}_{i}^{k}$. And it was proved that
the prediction error $\hat{e}_{i}^{k}$ is symmetric about zero,
which is independent of the choice of fitting order $p^{\prime}$
\cite{dufour nonparametric}. With this fact, $\hat{e}_{i}^{k}$ and
$-\hat{e}_{i}^{k}$ shall share the same distribution so that their
probability $\Pr(e_{i}^{k}>0)=\Pr(-e_{i}^{k}>0)=\Pr
(e_{i}^{k}<0)=1/2$ (with the assumption that $\Pr(e_{i}^{k}=0)=0$
in the sense of Lebesgue measure). Furthermore, let
$I_{i}(e_{i}^{k})$ denote the indicator of $e_{i}^{k}$ satisfying
that $I_{i}(e_{i}^{k})=1$ if $e_{i}^{k}>0$ and
$I_{i}(e_{i}^{k})=-1$ otherwise, then it can be shown that
$\{I_{i}(e_{i}^{k})\}$ is an independent series \cite{luger
exact}.

The signed rank statistic $\emph{SR}_{m}$ can be derived from a
set of $m$ prediction errors $\{\hat{e}_{i}^{k}:
i=i_{s_1},...,i_{s_m}\}$, where $\{i_{s_j}: j=1,2,...,m\}$ are
admissible indices for a given time series. Without lost of
generality, let us suppose the time indices $i=1,2,...,m$, then a
test statistic can be constructed as follows
\begin{equation}
\emph{SR}_{m}=\sum\nolimits_{i=1}^{m}I_{i}(e_{i}^{k})\times
S_{i}(rank(|e_{i}^{k}|)) \text{,} \label{signed rank}
\end{equation}
where $\{S_{i}(rank(|e_{i}^{k}|)\}_{i=1}^m$ is the set of scores
of $\{|e_{i}^{k}|\}_{i=1}^m$ with $rank(|e_{i}^{k}|)$ denoting the
rank of the absolute value $|e_{i}^{k}|$ among the set
$\{|e_{i}^{k}|\}_{i=1}^{m}$. In this work we choose
$S_{i}(rank(|e_{i}^{k}|))=rank(|e_{i}^{k}|)$ so that the test
statistic is reduced to the Wilcoxon signed rank statistic
\cite{luger exact,luo exact}
\begin{equation}
\emph{W}_{m}=\sum\nolimits_{i=1}^{m}i\times I_{i}(e_{i}^{k})
\text{.} \label{wilcoxon}
\end{equation}
For more detail about the signed rank statistic, the readers are
referred to, for example, \cite{maritz distribution}.

In Eq.(\ref{wilcoxon}) $\emph{W}_{m}$ is discretely distributed.
Theoretically one can enumerate all of its possible values and
thus obtain the full knowledge of its distribution. But in
practice enumeration will be inefficient for large $m$. Thus for
the sake of convenience, it is suggested to use the Gaussian
distribution $N(0,m(m+1)(2m\,+1)/6)$ for approximation based on
the central limit theorem, as adopted in our work. It was shown
that, even for a small integer, say $m=6$, one can still
approximate the discrete distribution quite well \cite{maritz
distribution}.

The above deduction means that the Wilcoxon signed rank statistic
$\emph{W}_{m}$ is a pivotal test statistic for the linear stochastic
processes with jointly symmetric innovations. But for a general
linear stochastic process, the foregoing conclusion does not
necessarily hold because of the possible asymmetry of innovation
terms. Nevertheless, we may apply the linear superposition
principle, as will be described below, to reduce the asymmetry based
on the central limit theorem. Through numerical experiments we will
also show that, for nonlinear processes, the summation over its
realizations might exhibit different characters from those of linear
stochastic ones.

Without lost of generality, let us consider, for example, two
linear stochastic time series $\{x_s\}_{s=j}^{j+l}$ and
$\{x_t\}_{t=k}^{k+l}$ produced by the same process described in
Eq. (\ref{AR_P}), then their summation $\{y_h\}_{h=0}^l \equiv
\{y_h: y_h=c_1 x_{j+h}+c_2 x_{k+h}\}_{h=0}^l$, for any
coefficients $c_1$ and $c_2$, also follows a linear stochastic
process
\begin{equation}
y_{h}=(c_1+c_2)a_{0}+\sum\nolimits_{j=1}^{p}a_{j}y_{h-j}+(c_1
\epsilon _{s+h}+c_2 \epsilon _{t+h}) \label{NewAR}
\end{equation}
with the innovation terms being $(c_1 \epsilon _{s+h}+c_2 \epsilon
_{t+h})$. This conclusion can be extended to general situations
with, say, $n$ time series. And then the drift and the innovation
term become $(\sum_{i=1}^n c_i) a_0$ and $\sum_{i=1}^n c_i
\epsilon_{t_i}$, where $c_i$ and $t_i$ are the coefficient and
time index associated with the $i$-th time series respectively.

Since the white noise $\{ \epsilon_i \}$ under consideration is
considered as an independent \footnote{And even identically
distributed, which, however, is not a necessary condition for our
deduction} sequence with finite variance (because of the
stationarity), the central limit theorem is applicable to the
summation $\sum_{i=1}^n c_i \epsilon_{t_i}$, i.e., one may expect
that the distribution of $\sum_{i=1}^n c_i \epsilon_{t_i}$ can be
approximated by a normal distribution for large $n$. Therefore,
even if originally $\epsilon_{t_i}$ does not have a jointly
symmetric distribution, the asymmetry would be reduced by summing
over a number of independent variables. Thus the Wilcoxon
statistic $\emph{W}_{m}$ is an asymptotically pivotal measure for
the summation of a number of linear stochastic time series. In
general, depending on the choice of scores in Eq. (\ref{signed
rank}), one may derive a family of asymptotically pivotal test
statistics.

In the following we will verify the above idea through few
numerical examples, which are: $\left(a\right)$ $ARMA(1,1)$
process $x_i = 0.1+0.5x_{i-1}+\epsilon_i-0.5 \epsilon_{i-1}$,
where $\epsilon_i$ follows the normal distribution $N(0,1)$;
$\left(b\right)$ $AR(2)$ process $x_i =
0.9x_{i-1}-0.3x_{i-2}+\eta_i$ with $\eta_i$ following the
(asymmetric) beta distribution $f(x)=x(1-x)^4/B(2,5)$, where
$B(2,5)$ denotes the beta function. For comparison of the
performance, we also include two nonlinear cases: $\left(c\right)$
Henon map $H(x,y)=(y+1-1.4x^{2},0.3x)$; and $\left(d\right)$
R\"{o}ssler system
$(\dot{x},\dot{y},\dot{z})=(-y-z,x+0.15y,0.2+xz-10z)$, where for
both cases the $x$ components will be singled out for calculation.

\begin{figure}[!t]
\centering
\includegraphics[width=3in]{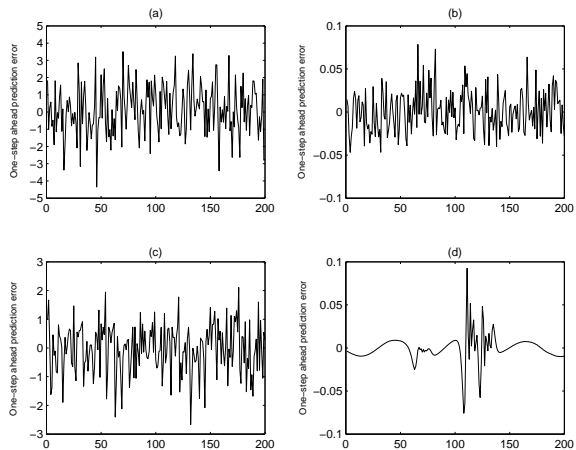}
\caption{Prediction errors ($p^\prime=8$) of: (a) $ARMA(1,1)$
process, (b) $AR(2)$ process, (c) Henon map, and (d) R\"{o}ssler
system, which are obtained by directly applying the OLS predictor
to the time series without summing over different realizations.}
\label{test_noAddition}
\end{figure}

\begin{figure}[!t]
\centering
\includegraphics[width=3in]{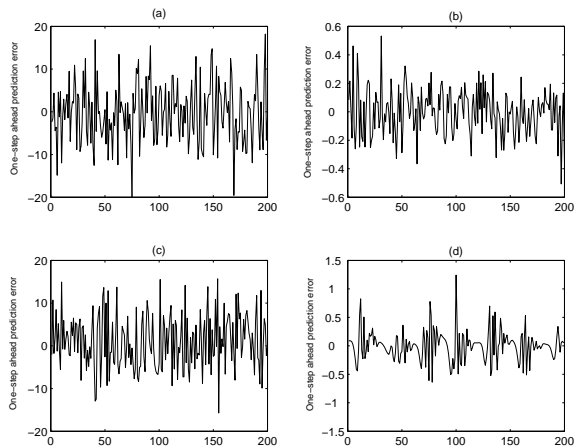}
\caption{Prediction errors ($p^\prime=8$) of summations over 20
realizations of: (a) $ARMA(1,1)$ process, (b) $AR(2)$ process, (c)
Henon map, and (d) R\"{o}ssler system.} \label{test_Addition}
\end{figure}

As the first step, we apply the Wilcoxon statistic $\emph{W}_{m}$
to directly examine the above data generation processes (DGPs). At
this stage, let us test the \emph{null hypothesis} which assumes
that the time series under examination is linear stochastic with
jointly symmetric innovations, while later we will extend the
hypothesis to a broader range as aforementioned.

For each DGP, the hypothesis testing is conducted 1000 times. Each
time a different realization of the same DGP is produced by varying
the initial condition(s). Each realization contains 2000 data
points, among which the last 200 points are taken out for
out-of-sample one-step ahead prediction based on the OLS predictor
in Eq. (\ref{OLS}), thus the number of prediction errors $m$=200.
Recall that, for a linear stochastic process with symmetric
innovation terms, theoretically the fitting order $p^\prime$ in Eq.
(\ref{OLS}) does not affect the symmetry of the prediction errors,
so in our calculations we do not choose any particular fitting
orders, instead we simply set $p^\prime=6,7,8,9,10$ for all cases.
For demonstration, one example of such prediction errors is plotted
in Fig. \ref{test_noAddition} for each DGP ($p^\prime$=8).

\begin{table}[!!t]
\centering
 \caption{\label{table1} Rejections of the null
hypothesis for $1000$ realizations of each DGP with different
fitting orders.}
\begin{tabular*}{2.5in}{ccccccccc}
\hline\hline
& & & DGP & & & Rejections of fitting order & &\\
\cline{6-8}
\end{tabular*}
\begin{tabular*}{2.5in}{ccccccccc}
& & & 6 & 7 & 8 & 9 & 10 \\ \hline
ARMA(1,1)& & & 56 & 55 & 53 & 52 & 53 \\
AR(2) & & & 78 & 83 & 80 & 80 & 81 \\
Henon & & & 73 & 64 & 69 & 61 & 67 \\
R\"{o}ssler & & & 79 & 396 & 920 & 862 & 963 \\
\hline\hline
\end{tabular*}
\end{table}

\begin{table}[!t]
\centering
 \caption{\label{table2} Rejections of the null
hypothesis for $1000$ summations over 20 realizations of each DGP
with different fitting orders.}
\begin{tabular*}{2.5in}{ccccccccc}
\hline\hline
& & & DGP & & & Rejections of fitting order & &\\
\cline{6-8}
\end{tabular*}
\begin{tabular*}{2.5in}{ccccccccc}
& & & 6 & 7 & 8 & 9 & 10 \\ \hline
ARMA(1,1)& & & 55 & 53 & 54 & 52 & 53 \\
AR(2) & & & 45 & 43 & 46 & 48 & 48 \\
Henon & & & 74 & 61 & 71 & 66 & 72 \\
R\"{o}ssler & & & 0 & 220 & 820 & 936 & 974 \\
\hline\hline
\end{tabular*}
\end{table}

Since the distribution of $\emph{W}_{m}$ is known
($N(0,m(m+1)(2m\,+1)/6)$), we can specify the false rejection rate
$\alpha$ (i.e., the rate that one falsely rejects the null
hypothesis), which is associated with two critical points
$C_{\alpha /2}$ and $C_{1-\alpha /2}$ for two-sided tests, where
in general notation $C_\delta$ denotes the critical point at which
the probability function $Pr(x)$ of a random variable $x$
satisfies that $Pr(x \leq C_\delta)=\delta$. Therefore if the
value of $\emph{W}_{m}$ in a test falls on the interval
$\left[C_{\alpha /2}, C_{1-\alpha /2}\right]$, we do not reject
the hypothesis, otherwise we reject.

In our calculations, we set $\alpha=5\%$ for all cases. Therefore
one would expect that, for a process consistent with the null
hypothesis, e.g., the $ARMA(1,1)$ process, the actual rejection
rate shall be the same as the nominal one $5\%$ (with slight
statistical fluctuations in practice).

But for the processes not consistent with the null, this
conclusion is not necessarily true. In fact, as indicated in Table
\ref{table1}, the actual rejection rates of the $AR(2)$ process,
the Henon map and the R\"{o}ssler system deviate from the nominal
in all cases, therefore we could reject the null hypothesis, which
means that those three DGPs cannot be linear stochastic processes
with jointly symmetric innovation terms.

Next let us extend the tests to a broader range, i.e., we consider
the \emph{null hypothesis} assuming that the time series is linear
stochastic governed by Eq. (\ref{AR_P}), but now let us remove the
assumption of joint symmetry previously imposed on the innovation
term $\epsilon_i$. In general, the prediction errors are not
necessarily symmetric, consequently the Wilcoxon statistic might
not really follow the expected distribution, thus the actual
rejection rates would deviate from the specified one. For example,
see the $AR(2)$ process (with the innovation term following the
asymmetric beta distribution) demonstrated in Table \ref{table1}.

The above problem could be tackled by instead testing the
summations over a number of different realizations of the same
underlying process, as discussed previously. In our calculations,
20 realizations are summed over for each test (As before, there
are 1000 tests in total). Before the summation each of them is
multiplied by a random coefficient uniformly distributed on
$[1,2]$, therefore all of the realizations are nearly equally
weighted to prevent over-dominance of any particular sequence. For
comparison, all the other computation settings are simply the same
as those adopted at the first step, and the test results are
reported in Table \ref{table2}, from which we could see that,
statistically the tests on the summations do not affect the
rejection rate of the $ARMA(1,1)$ process. However, for the
$AR(2)$ process its rejection rates now become very close to the
nominal one (with possible statistical fluctuations), which
supports our argument. In contrast, the rejection rates of the
Henon map and the R\"{o}ssler system still deviate from the
expected one, therefore one could reject the null hypothesis,
which means that their realizations are not generated by linear
stochastic processes. Again, for demonstration we plot in Fig.
\ref{test_Addition} the prediction errors of the summations for
each DGP ($p^\prime=8$).

Up to now, we have presented two folds of advantages of the Wilcoxon
signed rank statistic derived in this work. One is that, this
measure has an explicitly enumerable distribution, therefore one
could specify the exact false rejection rate for surrogate testing,
which is an unreachable property for many test statistics adopted in
the relevant literature. The other is that, this statistic is
pivotal for a special family of linear stochastic processes,
including the widely studied linear Gaussian noise. Moreover, it
also exhibits to be asymptotically pivotal for the summations of
realizations of general linear stochastic processes. Because of this
property it becomes possible now to extend the null hypothesis of
surrogate testing to more general situations.

\end{document}